\documentclass[aps,prl,groupedaddress,superscriptaddress,twocolumn,10pt,longbibliography]{revtex4-2}
\usepackage[utf8]{inputenc}
\usepackage{hyperref}
\usepackage{graphicx}
\graphicspath{{../figures/}}
\usepackage{setspace}
\usepackage{orcidlink}
\newcommand{\corauthor}[2]{
    \author{#1}
    \email{#2}
}

\usepackage{braket}
\usepackage{amsmath}
\usepackage{amssymb}
\begin{document}

\title{Competing Interactions in Strongly Driven Multi-Level Systems}
\corauthor{Jana Bender\orcidlink{0000-0002-1410-7330}}{agott-publication@physik.rptu.de}
\affiliation{Department of Physics and Research Center OPTIMAS, Rheinland-Pfälzische Technische Universität Kaiserslautern-Landau, 67663 Kaiserslautern, Germany}

\author{Patrick Mischke\orcidlink{0000-0001-7859-8426}}
\affiliation{Department of Physics and Research Center OPTIMAS, Rheinland-Pfälzische Technische Universität Kaiserslautern-Landau, 67663 Kaiserslautern, Germany}
\affiliation{Max Planck Graduate Center with the Johannes Gutenberg-Universität Mainz (MPGC), Staudinger Weg 9, 55128 Mainz, Germany}

\author{Tanita Klas\orcidlink{0000-0001-7638-0490}}
\affiliation{Department of Physics and Research Center OPTIMAS, Rheinland-Pfälzische Technische Universität Kaiserslautern-Landau, 67663 Kaiserslautern, Germany}

\author{Florian Binoth\orcidlink{0009-0004-3024-6412}}
\affiliation{Department of Physics and Research Center OPTIMAS, Rheinland-Pfälzische Technische Universität Kaiserslautern-Landau, 67663 Kaiserslautern, Germany}

\author{Hani Naim}
\affiliation{Department of Physics and Research Center OPTIMAS, Rheinland-Pfälzische Technische Universität Kaiserslautern-Landau, 67663 Kaiserslautern, Germany}

\author{Herwig Ott\orcidlink{0000-0002-3155-2719}}
\affiliation{Department of Physics and Research Center OPTIMAS, Rheinland-Pfälzische Technische Universität Kaiserslautern-Landau, 67663 Kaiserslautern, Germany}

\author{Thomas Niederprüm\orcidlink{0000-0001-8336-4667}}
\affiliation{Department of Physics and Research Center OPTIMAS, Rheinland-Pfälzische Technische Universität Kaiserslautern-Landau, 67663 Kaiserslautern, Germany}

\date{\today}

\begin{abstract}
We experimentally study the level mixing, splitting and repulsion of an optically driven atomic multi-level system under two competing interactions.
The strength of the optical coupling is increased until it surpasses the atomic hyperfine interaction responsible for mixing the magnetic substates.
Due to the multi-level character of the coupled state space, the level shifts exhibit complex behavior reminiscent of the Paschen-Back effect.
Our results show that multi-level effects can have significant influence for strong external drive, differing from a simple model of effective non-interacting two-level systems.
These results highlight the relevance of imperfections of the light polarization or initial state preparation in strongly optically driven systems.

\end{abstract}

\maketitle


\section{Introduction}

A detailed understanding of the interaction between atoms and electromagnetic fields is a cornerstone of modern atomic physics.
The high level of control of magnetic and optical transitions has enabled groundbreaking precision applications such as optical clocks \cite{Ludlow2015} and interferometers \cite{Bongs2019}.
It is also key for applications in quantum information and quantum computing with atomic qubits \cite{Bruzewicz2019,Saffman2016}.

While the multi-level structure of a real quantum system can provide interesting features \cite{brion_quantum_2007, zhou_full_2023}, in many cases, it is rather undesirable and one prefers to work with a two-level sub system such that the experimental implementation and the theoretical description is simplified.
If necessary, an interaction with other levels can then be included via perturbation theory.
The situation becomes complex, however, if the interaction energy with the electromagnetic field is comparable to that of an internal coupling mechanism within the system.
As a consequence, more than two levels are involved in the dynamics, and mixing of levels as well as non-trivial line shifts take place.
A well known example for this situation is the Paschen-Back effect, where the interaction with a static magnetic field competes with the hyperfine interaction of an atom \cite{paschenback_1921}.
Also DC Stark effects in quantum defect Rydberg systems show a similar phenomenology \cite{zimmerman_stark_1979, zhang_stark-induced_2013} and, more generally, similar situations occur for strong AC electromagnetic field drive (RF, microwave or optical frequencies) competing with an internal atomic (or molecular) interaction mechanism.\\
Due to their strong an long-range interaction \cite{Adams2020,Browaeys2020}, atoms in Rydberg states are currently under intense investigation and present themselves as promising candidates for atom-based information processing protocols\cite{Saffman2016}.
Frequently the excitation of Rydberg states relies on a two-photon transition, where the upper transition is driven strongly.
The quest for ever larger Rabi frequencies to the Rydberg state often entails a situation where the Rabi coupling becomes comparable to the hyperfine interaction of the intermediate state or the fine structure interaction in the Rydberg state.
To understand the spectroscopical measurements on multi-level Rydberg systems presented here, it is thus necessary to properly account for the very generic phenomenon of multi-level effects in such strongly driven systems.


\begin{figure}
    \hspace{-1cm}\includegraphics[scale=1]{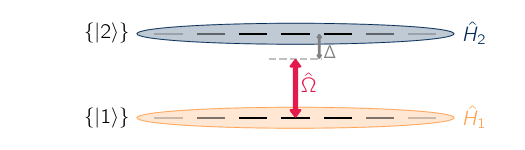}
    \caption{Two generalized sets of energy levels coupled with interaction strengths $\Omega_{i, k}$ and energy offset $\Delta$.
    The states $\{\ket{1}\}$ in the lower set have an interaction Hamiltonian $\hat{H}_1$, the states $\{\ket{2}\}$ in the upper set interact via $\hat{H}_2$.}
    \label{fig:level_sketch}
\end{figure}

One of the most basic and powerful models to describe quantum phenomena is a system of two states $\ket{1}$ and $\ket{2}$ which are coupled to each other by an external interaction  $\hat{H}_\mathrm{ext} = \frac{\hbar\Omega}{2}\hat{\sigma}_x$.
With the energy offset $\Delta = E_2 - E_1$ the Hamiltonian for such a system reads in rotating wave approximation ($\hbar = 1$)
\begin{equation}
    \hat{H} = \begin{pmatrix}
        0                & \frac{\Omega}{2} \\
        \frac{\Omega}{2} & \Delta
    \end{pmatrix}.
    \label{eq:twolevelhamilton}
\end{equation}
Diagonalization of the above Hamiltonian provides two dressed eigenstates $\ket{\pm} = \frac{1}{\sqrt{2}}\left(\ket{1} \pm \ket{2}\right)$, whose energies are given by $E_\pm = \pm\frac{\hbar \Omega}{2}$ in the limit  $\Omega \ll \Delta$.\\
While this simple model allows for valuable insights, real physical systems typically feature a multi-level structure, which it cannot accurately describe.
A straight-forward extension, which is encountered frequently in atomic system, is to consider two sets $\{\ket{1}\}$ and $\{\ket{2}\}$ of levels, which are coupled to each other (see \,Fig.\,\ref{fig:level_sketch}).
Analogous to \,Eq.\,\ref{eq:twolevelhamilton}, the Hamiltonian for such a system can be written as ($\hbar=1$)
\begin{equation}
    \hat{H} = \begin{pmatrix}
        \hat{H}_\mathrm{1} & \hat{\Omega}^{\dagger}/2    \\
        \hat{\Omega}/2     & \hat{H}_\mathrm{2} + \Delta \mathbb{I}
    \end{pmatrix}
    \label{eq:multihamilton}
\end{equation}
where $\hat{\Omega}$ denotes the couplings between the sets of states with detuning $\Delta$ and $\hat{H}_1$ and $\hat{H}_2$ denote internal interactions within each subset.
For degenerate subsets, one can apply the Morris-Shore transformation \cite{morris_reduction_1983} and perform a basis transformation, decoupling the Hilbert space into independent two-level systems and non-participating dark states.
However, real quantum systems frequently feature internal interactions like spin-orbit or spin-spin couplings.
In such a situation the internal interactions $\hat{H_i}$ couple the states within the respective subset and the degeneracy is lifted. The Morris-Shore transformation can then no longer be applied exactly.
The composition and the energies of the eigenstates and the treatment of the system now strongly depend on the relative magnitude of the two energy scales imposed by $\hat{H}_i$ and $\hat{H}_\mathrm{ext}$. Only in the limit $\hat{H}_i \ll \hat{H}_\mathrm{ext}$, an approximate eigenvalue approach remains possible \cite{zlatanov_morris-shore_2020}.
In the general case, the internal interaction mechanism results in multi-level subsystems of various sizes whose eigenstates can only be obtained by a full diagonalization.
When the strength of the external interaction is increased, the system will eventually change from being dominated by $\hat{H}_\mathrm{i}$ to being dominated by $\hat{H}_\mathrm{ext}$.
This is the basic mechanism behind effects like the DC Stark effect in magnetic fields and the Paschen-Back effect. 
For $\hat{H}_i \approx \hat{H}_\mathrm{ext}$ the system features strong level mixing.\\
Theoretical investigations have been conducted in particular with focus on the interplay of the hyperfine interaction of atomic levels with light field coupling \cite{bevilacqua_bright_2022, kirova_hyperfine_2017}.
While those studies focused on the limiting case of weak hyperfine interaction (and which can therefore still be treated by the Morris-Shore transformation), we pay attention to the crossover regime, where $\hat{H}_i$ and $\hat{H}_\mathrm {ext}$ are comparable.

\section{Model}

\begin{figure}
    \includegraphics[width=.49\textwidth]{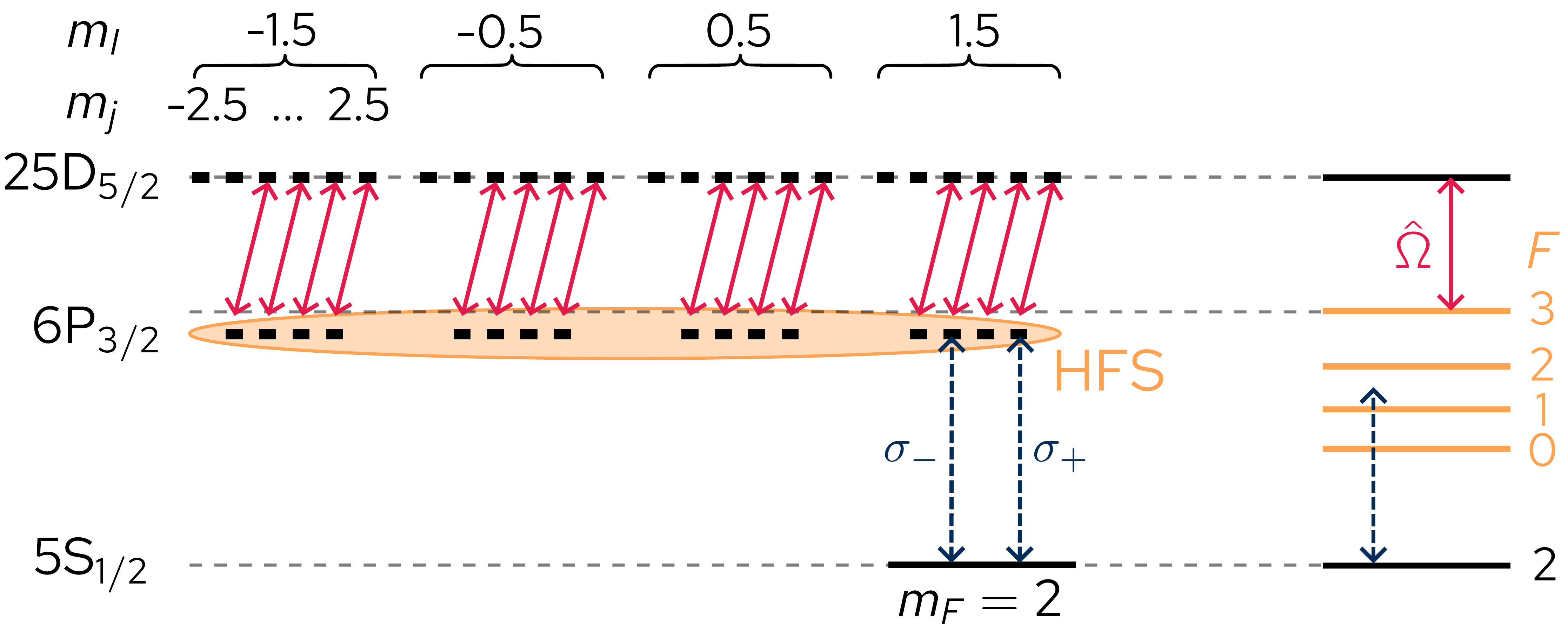}
    \caption{Multi-level system in rubidium 87 with the magnetic substates of $\{\ket{6\mathrm{P}_{3/2}}\}$ and $\{\ket{25\mathrm{D}_{5/2}}\}$ under $\sigma^+$ coupling.
        On the left, the levels are depicted in the fine structure basis.
        On the right, the energy level sets of the hyperfine basis states are sketched.
        The coupling light (red arrows) is resonant to the transitions starting from the $\{\ket{6\mathrm{P}_{3/2}, F=3, m_F}\}$ hyperfine states.
        A second, weak laser probes the system from the $\ket{5\mathrm{S}_{1/2}, m_j=0.5, m_I=1.5}=\ket{5\mathrm{S}_{1/2}, F=2, m_F=2}$ ground state (blue arrows).
        It can be detuned over the full $\{\ket{6\mathrm{P}_{3/2}}\}$ hyperfine manifold.
        Depending on its polarization, it either couples to $\ket{6\mathrm{P}_{3/2}, m_j=1.5, m_I=1.5}$ ($\sigma^+$ transition) or $\ket{6\mathrm{P}_{3/2}, m_j=-0.5, m_I=1.5}$ ($\sigma^-$ transition).
    }
    \label{fig:rb_level_scheme}
\end{figure}

Here, we study an instance of this scheme, which is realized by the strong optical coupling in a multi-level atomic system. Specifically, we consider the transition in rubidium from the $6P_{3/2}$  state to the Rydberg $25D_{5/2}$ state.
We explicitly take into account the  magnetic sublevels of both manifolds as well as the hyperfine interaction of the $\ket{6\mathrm{P}_{3/2}}$, see \,Fig.\,\ref{fig:rb_level_scheme}.
Note that the hyperfine interaction of the Rydberg state can be neglected. The resulting line shifts and level mixing are weakly probed by optical spectroscopy from the $5\mathrm{S}_{1/2}$ ground state.

%
%
\begin{figure*}
    \center
    \includegraphics[scale=1]{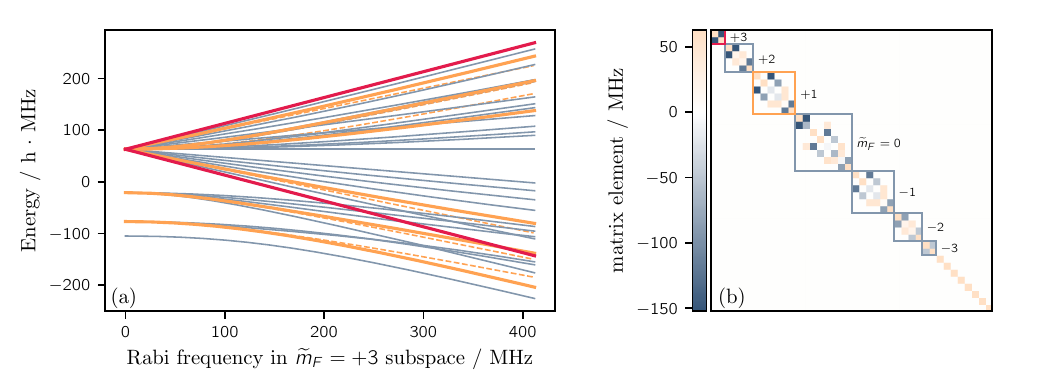}
    \caption{
        (a) The eigenenergies of the Hamiltonian \,Eq.\,\ref{eq:fullhamilton} (blue).
        Zero energy is chosen to coincide with the energy of the uncoupled $\{\ket{6\mathrm{P}_{3/2}, m_j, m_I}\}$ states.
        The red and orange lines correspond to the eigenenergies of the two subsystems highlighted in (b).
        The dashed orange lines indicate extrapolations of the two-level behavior in the weak coupling limit (see text).
        (b) Graphical representation of the eigenstates of the Hamiltonian \,Eq.\,\ref{eq:fullhamilton} for pure $\sigma^+$ coupling. The states have been reordered, such that the block diagonal form is visible (see text). The Rabi frequency in the $\widetilde{m}_F = +3 $ subspace is set to 200\,MHz.
        The system decouples into independent subsystems of different size. The blocks are labeled by $\widetilde{m}_F = +3,2, .. , -3$ followed by uncoupled individual states in the lower right corner. The red box highlights the $\widetilde{m}_F = +3$ subspace that forms a two-level system.
        The orange box highlights $\widetilde{m}_F=+1$ subspace that forms a six-level system.}
    \label{fig:two_example_systems}
\end{figure*}
We formulate the problem in the uncoupled hyperfine structure basis spanned by the 16 possible $\{\ket{6\mathrm{P}_{3/2}, m_j, m_I}\}$ states and the 24 possible $\{\ket{25\mathrm{D}_{5/2}, m_j, m_I}\}$.
The latter are all degenerate.
Setting up \,Eq.\,\ref{eq:multihamilton} and explicitly adding the hyperfine interaction in the lower manifold yields ($\hbar=1$)
\begin{equation}
    \hat{H} = \begin{pmatrix}
        \hat{H}_{6\mathrm{P}} & \hat{\Omega}^{\dagger}/2 \\
        \hat{\Omega}/2        & \Delta \mathbb{I}
    \end{pmatrix}
    \label{eq:fullhamilton}
\end{equation}
with the subspace Hamiltonians $\hat{H_1}=\hat{H}_{6\mathrm{P}} = A_\mathrm{HFS} \mathbf{I \cdot J}$ and $\hat{H_2}=0$. The matrix $\hat{\Omega}$ summarizes all couplings between the $\{\ket{6\mathrm{P}_{3/2}, m_j, m_I}\}$ and the $\{\ket{25\mathrm{D}_{5/2}, m_j, m_I}\}$ states.
The coupling laser is resonant with the hyperfine states $\{\ket{6\mathrm{P}_{3/2},F=3, m_F}\}$.
The Hamiltonian \,Eq.\,\ref{eq:fullhamilton} is then diagonalized.
\,Fig.\,\ref{fig:two_example_systems}a) shows the resulting eigenenergies in dependence of the coupling strength $\Omega$.\\
In the limit of small coupling $\Omega \rightarrow 0$, we recover the hyperfine basis states $\{\ket{6\mathrm{P}_{3/2}, F, m_F}\}$ consisting of four sets of degenerate states with $0 \leq F \leq 3$.
In this limit, the AC stark shift can be regarded as a perturbation.
As predicted by the Morris-Shore transformation, the resulting eigenenergy splitting corresponds to the behavior of independent two-level systems (resonantly and off-resonantly driven), which emerge from the hyperfine manifolds.\\
In the limit of strong coupling (not shown in the plot), the eigenstates can again be grouped into two-level systems whose energies split symmetrically and linearly, with the hyperfine interaction as a perturbation.\\
We can identify two different kinds of eigenstates \cite{kirova_hyperfine_2017}.
On the one hand there are dark states in the $\{\ket{25\mathrm{D}_{5/2}, m_j, m_I}\}$ subset, which are not coupled to any of the $\{\ket{6\mathrm{P}_{3/2}, m_j, m_I}\}$ states. Their eigenenergies show no shift at all.
On the other hand, there are so-called bright states experiencing an AC stark shift.\\

In between those limiting cases, the eigenenergies show a rich multi-level behavior (see \,Fig.\,\ref{fig:two_example_systems}).
As the Hamiltonian preserves the azimuthal symmetry, the projection of the total angular momentum of the atom-light system $\widetilde{m}_F = m_j + m_I  - n_p$ is conserved.
Here, $n_p$ is the angular momentum of the photon. For $\sigma^+$ transitions, it is 1 for the $\{\ket{25\mathrm{D}_{5/2}, m_j, m_I}\}$ and 0 for the $\{\ket{6\mathrm{P}_{3/2}, m_j, m_I}\}$ states.
The Hilbert space can then be separated in independent subspaces with fixed $\widetilde{m}_F$.
By reordering the basis states through the Reverse Cuthill McKee algorithm \cite{cuthill_reducing_1969}, these independent subsystems of different size become apparent in the block diagonal structure of the coupling matrix.
This is illustrated in the right part of \,Fig.\,\ref{fig:two_example_systems}. The plot shows the couplings between all states in the reordered hyperfine structure basis. The coupling strength is encoded in the color code.
The subsystem highlighted with a red box denotes the \textit{fully stretched} states with $\widetilde{m}_F = +3$, which are eigenstates of both the hyperfine and the finestructure basis.
The $\widetilde{m}_F = +3$ subspace forms an isolated two-level system and thus its eigenenergies show a symmetrical, linear Autler-Townes splitting unaffected by the internal interaction (red lines in Fig.\,\ref{fig:two_example_systems}a).
By setting the laser polarization to a $\sigma _+$-transition, this two-level subsystem allows to solely couple the $\ket{6\mathrm{P}_{3/2}, F=3, m_f=3}$ to the $\ket{25\mathrm{D}_{5/2}, m_j = 5/2, m_I = 3/2}$.
However, the light field coupling those states also couples other substates which fulfill the selection rules $\Delta m_j = 1$ and $\Delta m_I = 0$, leading to couplings in further subspaces.
An example of a more complex subspace with six states involved, characterized by $\widetilde{m}_F = +1$, is indicated by the orange box in Fig.\,\ref{fig:two_example_systems}b).
In the hyperfine basis, the states spanning this subspace correspond to the $\{\ket{6\mathrm{P}_{3/2}}\}$ states of variable $F$ with $m_F = 1$ which are coupled to the $\{\ket{25\mathrm{D}_{5/2}}\}$ states with $m_F = 2$.
In contrast to the simple two-level system, the eigenenergies in this subspace show a nonlinear, asymmetrical behavior in their splitting.
For comparison, the dashed orange lines extrapolate the behavior in the limit of small coupling where the AC stark shift can be treated as a small perturbation and behaves as in a two-level system:
One pair of states, originating from the energy of $\ket{6\mathrm{P}_{3/2}, F=3}$, is coupled resonantly and would show a linear splitting.
The others are coupled with a detuning according to the hyperfine splitting but still split up symmetrically.
In the diagonalization results, however, the six eigenenergies in the $\widetilde{m}_F=+1$ subspace repel each other and therefore deviate from this behavior.
The difference between the solid and dashed orange lines therefore shows the influence of the level mixing due to the strong coupling.\\

\section{Experiment}
\begin{figure*}
    \hspace{-1cm}\includegraphics[scale=1]{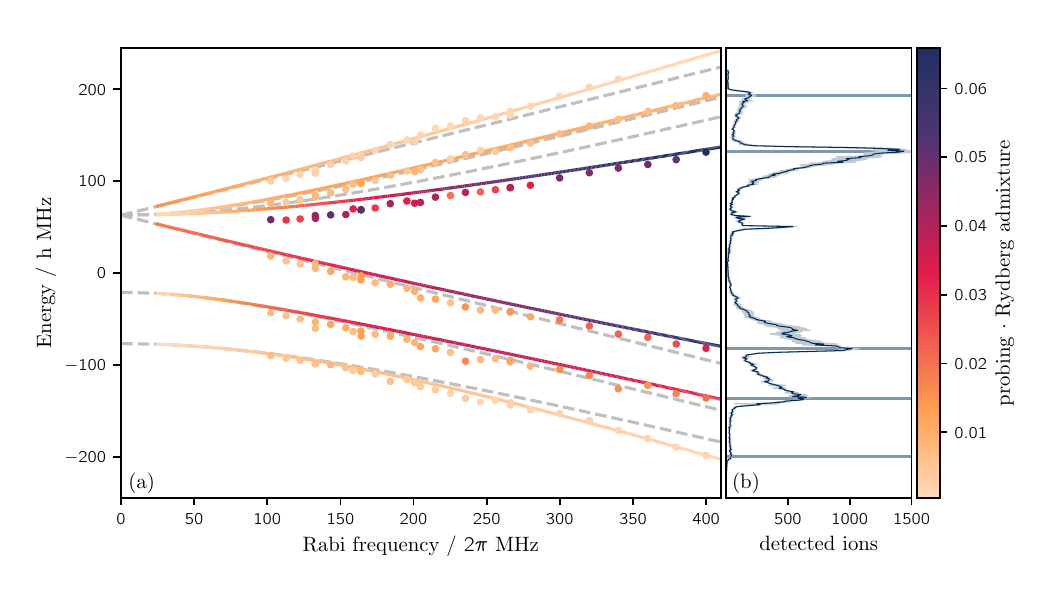}
    \caption{(a) Comparison of the experimental results (dots) with the calculated eigenenergies (lines) for the $\widetilde{m}_F=+1$ subsystem highlighted in \, Fig.\,\ref{fig:two_example_systems}.
        The color code of the theory lines gives a prediction for the expected experimental signal strength.
        The energies of the experimental peaks were fitted from measured ion spectra and the peak height is normalized globally and is represented in the dot colors.
        The dashed lines denote the expected peak positions for hypothetical independent two-level systems.
        (b) Exemplary measured spectrum at a Rabi frequency of 400 MHz with five peaks of the probed multi-level system.
        The peaks show an asymmetric broadening towards the center because of the inhomogeneous power distribution of the coupling laser.
        The solid lines represent the peak positions as determined with the fit.}
    \label{fig:exp_results}
\end{figure*}
We prepare a dilute sample of $\approx 5 \times 10^5$  rubidium 87 atoms in a crossed dipole trap in the $\ket{5\mathrm{S}_{1/2}, F=2, m_F=2}$ ground state.
The dipole trap beams intersect under a $45 ^{\circ} $ angle.
One of them is circularly polarized and acts as the coupling laser ($\lambda=1030\,$nm) driving $\sigma^+$-transitions.
The laser is resonant to the fully stretched transition $\ket{6\mathrm{P}_{3/2}, F=3, m_F=3} \rightarrow \ket{25\mathrm{D}_{5/2}, m_j = 5/2, m_I=3/2}$.
When varying the power of this beam in order to change the coupling strength, the atom number is kept constant by adapting the loading parameters of the trap.
The eigenstates resulting from the strong coupling are probed by a weak laser on the $\ket{5\mathrm{S}_{1/2}, F=2, m_F=2} \rightarrow \{\ket{6\mathrm{P}_{3/2}}\}$ transition with a wavelength of 420\,nm. The probe beam is counter-propagating to the coupling beam.
To maintain a constant trapping potential, the coupling light is turned on continuously during the experiment and only the probing laser is turned on for 1000 pulses of $2\,\mathrm{\mu s}$ length each.
The probe laser is circularly polarized and can drive $\sigma^+$ or $\sigma^-$ transitions. This way, we can probe different admixtures of the eigenstates generated by the coupling laser:
For $\sigma^+$ polarization, we probe the admixture of the  $\ket{6\mathrm{P}_{3/2}, F=3, m_F=3}$ state, while for $\sigma^-$ polarization we probe all states with an admixture of the $\ket{6\mathrm{P}_{3/2}, m_j=-1/2, m_I=3/2}$ state (corresponding  to all $\{\ket{6P_{3/2}}\}$ states with $m_F = 1$ but varying $F$).
This means that by driving a $\sigma^+$-transition, the $\widetilde{m}_F=3$ system highlighted in red in Fig.\,\ref{fig:two_example_systems}a) is probed while for a $\sigma^-$ probing, the six-level $\widetilde{m}_F = +1$ system highlighted in orange is probed.
It is crucial to have well-polarized beams because an admixture of $\sigma^-$ polarization in the coupling beam would result in couplings between the otherwise separate subsystems which complicates the identification of the peaks in the spectroscopy.
Similarly, a clean probe laser polarization is required to selectively do spectroscopy on only a single subsystem.
The coupling Rabi frequency is calibrated through measuring the Autler-Townes splitting of the $\widetilde{m}_F = +3$ two-level system in a separate measurement.\\
Due to their admixture of $\{\ket{25\mathrm{D}_{5/2}, m_j, m_I}\}$ Rydberg states, the probed eigenstates spontaneously decay into ions through photoionization by the trapping light or blackbody radiation.
These ions are extracted with a small electric field and guided to an ion detector which provides the primary measurement signal.
It allows for the detection of the target state with a very low background signal.
By continuously detuning the probe laser while recording the ion signal, we spectroscopically determine the eigenenergies of the coupled system.

\section{Results}

We are particularly interested in the six-level subsystem with $\widetilde{m}_F=+1$, where multi-level effects are pronounced.
We measure the energy spectrum for different values of the coupling strength.
By extracting the peak positions and amplitudes by Gaussian fits for every spectrum, we map out the eigenenergies for this subspace.
The results are plotted in \,Fig.\,\ref{fig:exp_results} a) where the dots represent the measured peak positions and the solid lines show the results of the diagonalization.
An exemplary spectrum is shown in \,Fig.\,\ref{fig:exp_results} b).
The asymmetric peak broadening towards the center of the spectrum originates from the different coupling Rabi frequency the atoms experience depending on their position in the crossed dipole trap.
A numerical simulation integrating over the spatial distribution of the coupling strength agrees well with the measured line shapes.\\
The measured eigenenergies show a very good agreement with the calculations which include no free parameter.
The multi-level character of this subsystem becomes apparent when comparing the measured data with the dashed lines.
As discussed above (see also \,Fig.\,\ref{fig:two_example_systems}a), they show the behavior of hypothetical independent two-level systems originating from the same asymptotic energy.
One can see that the state repulsion in the multi-level system increases the AC Stark shift for the outermost states and decreases it for the inner ones.
Especially the first and fourth state from above deviate from the simple splitting a resonantly coupled two-level system would exhibit.
The third state from above, as expected from the diagonalization, shows the largest absolute deviation from a simple two-level system.

The fit of the measured peak amplitudes also allows us to compare the peak heights with the expected coupling strength from the model calculations.
The color code of the solid lines in \,Fig.\,\ref{fig:exp_results}a) represents the theory prediction, given by
\begin{equation}
    \begin{split}
        p = &|\braket{6\mathrm{P}_{3/2}, m_j=-0.5, m_I=1.5|\Psi}|^2\\
        &\cdot \sum_{ m_j, m_I} |\braket{25\mathrm{D}_{5/2}, m_j, m_I|\Psi}|^2.
    \end{split}
    \label{eq:signal_prediction}
\end{equation}
It takes into account the projection onto the $\ket{6\mathrm{P}_{3/2}, m_j=-0.5, m_I=1.5}$ state coupled by the probe laser as well as the admixture of the ion-creating Rydberg states.
The experimentally extracted peak heights are globally normalized to the maximum of the theory prediction to allow a comparison between experiment and theory.
One can see that for stronger couplings the relative signal strengths agree qualitatively with the theory prediction.
The peak heights of the two innermost eigenstates fit the trend of a rising signal strength for stronger coupling.
Overall, the agreement between experiment and theory is very good.

Furthermore, the model predicts that the second lowest eigenstate shows a decreasing peak height for a Rabi frequency above 300 MHz (referring to the $\widetilde{m}_F = +3 $ subspace).
This is a characteristic of a so-called \textit{chameleon state}.
Such a state is characterized by a decreasing admixture of the probed state with increased coupling, resulting in its disappearance for strong coupling \cite{kirova_hyperfine_2017, cinins_expressions_2022}.
While the model clearly indicates the existence of a chameleon state in our system, the spread in the measured peak heights is too high to confirm it in the experimental data.\\

In summary, we have investigated a system of hyperfine states of rubidium 87 coupled to a Rydberg state by a strong light field.
The diagonalization yields a series of independent multi-level systems, whose structure ordered with respect to the  angular momentum projection quantum number.
We experimentally confirmed the predicted splitting of the $\widetilde{m}_F=+1$ subsystem and its deviation from the behavior of independent two-level systems.

\section{Conclusion and Outlook}

Our findings deepen the understanding of multi-level systems coupled by two competing interaction mechanisms.
In particular, we cast light on the influence of the spin-spin interaction under atom-light coupling.
To selectively couple a two-level system, both the states and the light field polarization have to be chosen carefully.
Otherwise, a multi-level system formed by additional interactions of the atomic levels is addressed.
In such a situation, the resulting AC stark shifts will no longer correspond to a series of independent two-level Autler-Townes splittings.
In the limits of very weak or strong Rabi coupling, when one interaction is dominating, a good theoretical modeling via perturbation theory is possible.
In between, a transitional regime exists where all couplings have to be considered to the same extent.\\
Our results demonstrate that multi-level effects become important, when working with atoms which are not cleanly prepared in one magnetic substate. The same applies for a non-perfect light polarization, which also enlarges the addressed subsystem.
A comprehensive understanding of the structure of the eigenstates not only helps to avoid unwanted couplings between the atomic transitions,
the multi-level structure might also be exploited, for example, by using it to prepare a desired target states by combining adiabatic and non-adiabatic ramps of the coupling light.


\section{Acknowledgements}
We thank Sebastian Hofferberth for very helpful discusssions on the spectral features.
We acknowledge financial support by the DFG within the collaborative research center TR-185 OSCAR (number 277625399) and within project number 460443971.
This work was also supported the research initiative Quantum Computing for Artificial Intelligence (QC-AI) and by the Max Planck Graduate Center MPGC with the University of Mainz.

%

\end{document}